\documentclass{elsart}
\usepackage{graphics,epsf}
\def\CM{{\mathcal M}}

\def\CP{{\mathcal P}}

\def\bK{\mbox{\boldmath $K$}}

\def\bR{\mbox{\boldmath $R$}}

\def\bk{\mbox{\boldmath $k$}}

\def\bp{\mbox{\boldmath $p$}}
\def\bq{\mbox{\boldmath $q$}}

\def\aagp{{\alpha \alpha^\prime}}

\def\eq#1{Eq.\,(\ref{#1})}

\begin{document}
\begin{frontmatter}


%
\title{Hyperon Single-Particle Potentials Calculated {\bf from}
$SU_6$ Quark-Model Baryon-Baryon Interactions}
\author{M. Kohno, Y. Fujiwara$^*$, T. Fujita$^*$,
 C. Nakamoto$^{**}$ Y. Suzuki$^{***}$}
\address{Physics Division, Kyushu Dental College,
Kitakyushu 803-8580, Japan \\
$\hbox{}^*$Department of Physics, Kyoto University, 
Japan \\
$\hbox{}^{**}$Suzuka National College of Technology,
Suzuka 510-0294, Japan \\
$\hbox{}^{***}$Department of Physics, Niigata University,
Niigata 950-2181, Japan}

\maketitle

\begin{abstract}
\hspace*{1em}
Using the $SU_6$ quark-model baryon-baryon interaction
recently developed by the Kyoto-Niigata group,
we calculate $NN$, $\Lambda N$ and $\Sigma N$ $G$-matrices
in ordinary nuclear matter. This is the first attempt to discuss the
\(\Lambda\) and \(\Sigma\) single-particle
potentials in nuclear medium, based on the realistic quark-model
potential. The \(\Lambda\) potential has the depth of more than $40$ MeV,
which is more attractive than the value expected from the
experimental data of $\Lambda$-hypernuclei.
The \(\Sigma\) potential turns out to be repulsive,
the origin of which is traced back to the strong Pauli repulsion
in the $\Sigma N (I=3/2)$ $\hbox{}^3S_1$ state.
\end{abstract}

\end{frontmatter}

\section{Introduction}
The $SU_6$ quark-model provides a unified framework to describe
baryon-baryon interactions including hyperons \cite{WS84,FU87}.
Recently the Kyoto-Niigata group has developed a modern quark-model
baryon-baryon interaction \cite{NA95,FU95,FU96a,FU96b,FJ96a,FJ96b,FJ98},
which reproduces essential features of the nucleon-nucleon ($NN$) and
hyperon-nucleon ($YN$) scattering data below 300 MeV.
In this model, the quark-quark interaction is assumed to consist of
a phenomenological quark-confining potential, the Fermi-Breit
interaction coming from the one-gluon exchange mechanism and
effective meson-exchange potentials
of scalar and pseudo-scalar meson nonets directly coupled to quarks.
The interaction between the baryons is then derived in a framework
of $(3q)$-$(3q)$ resonating-group method (RGM).
The essential difference between the traditional meson-exchange
potentials and the present quark model lies in a description of
the short-range part of the interaction.
In the traditional approach the baryon is treated as structureless
and the short-range repulsion
is introduced more or less phenomenologically
with the aid of a hard-core, $\omega$-meson exchange,
or more recently pomeron-exchange between baryons.
On the other hand, the compositeness of the baryon is
explicitly considered in the quark Hamiltonian,
and the origin of the short-range repulsion is the quark-exchange
kernel of the color-magnetic term contained in the Fermi-Breit
interaction, as well as the strong effect of the Pauli principle
acting in some specific channels.
Since the whole interaction acting between the baryons is
determined by a strong cancellation
between this repulsion at the short-range and the
intermediate-range attraction, it is natural to expect
that these two models predict quite different results
for some observables whose experimental data are still
not yet available.
In fact, we have shown in the previous
papers \cite{FJ96a,FJ96b,FJ98} that,
in some $YN$ observables and also in certain partial waves,
the quark-model potential gives predictions different
from that of the one-boson exchange model
such as the Nijmegen model \cite{HCD,HCF,NSC,NSC99}
and the J{\" u}lich model \cite{HO89,RE94}.
Since available experimental data are still scarce in the strangeness
sector, it is useful to elucidate further the characteristics of
the quark-model potential and to pursue its implications to
hypernuclear physics.

Although the \(\Lambda\) single-particle (s.p.) potential
in nuclei has been established experimentally \cite{BMZ},
the \(\Sigma\) potential is basically unknown and even the sign of
the \(\Sigma\) potential has been controversial \cite{BAT}.
Since there are no conclusive experimental data
of the $\Sigma$ hypernuclear states,
a theoretical estimation of the $\Sigma$ potential is made by
extending known or unknown coupling constants to the $\Sigma$ sector.
The Nijmegen model D and soft-core potentials suggest
an attractive $\Sigma$ s.p. potential \cite{YB85,YB90,SCHU},
while the Nijmegen model F indicates a repulsive one \cite{YB90}.
In a relativistic mean-field description \cite{RMF},
the results depend on the ratios of
the coupling constants $\alpha_i \equiv (g_{i \Sigma^0}/g_{i N})$
~($i= \sigma, \omega, \rho$) which is not known a priori.
Earlier studies \cite{GLEN}, assuming a universal coupling,
led to a $\Sigma$ potential
which is basically equal to the $\Lambda$ one.
A different choice of the ratios was found later \cite{MARE}
to be able to give a repulsive  $\Sigma$ potential.
In view of the fact that various models give different predictions,
it is interesting to discuss theoretical predictions of the
\(\Sigma\) potential obtained from a quark model which unifies
a description of $NN$ and $YN$ interactions.

In this paper we present $G$-matrix calculations,
in the lowest order Brueckner theory \cite{LOBT,DAY}, for the $NN$,
$\Lambda N$ and $\Sigma N$ interactions in ordinary nuclear matter,
by using the quark-model interaction developed by the
Kyoto-Niigata group. There are several versions of the quark model;
RGM-F \cite{NA95,FU95}, FSS \cite{FU96a,FU96b,FJ96a,FJ96b,FJ98} and
RGM-H \cite{FU96b,FJ96a,FJ96b,FJ98}.
We report mainly the results with the FSS, since it incorporates
the effective meson-exchange potentials in the complete microscopic way.
Although $\Lambda p$ total cross section in the cusp region
is somewhat overestimated due to the strong antisymmetric $LS$ force
($LS^{(-)}$ force), the threshold energy of the $\Sigma N$ channel
is correctly reproduced.
We discuss nuclear-matter saturation properties and s.p. potentials
of the $N$, $\Lambda$ and $\Sigma$ obtained from the $G$-matrices.
These properties predicted by the other versions
of our quark model, RGM-F and RGM-H, are essentially the same.
Further analysis of partial-wave contributions
enables us to clarify general characteristics of our quark model.

The problem of the strength of the hyperon s.p. spin-orbit
potential is another interesting and important issue.
The analysis on this subject based on the quark-model
$G$-matrices is treated in a separate paper \cite{LS99}.

We outline in Section 2 a calculation of $G$-matrices
from the quark-model baryon-baryon interactions.
The saturation property of nuclear matter, predicted
by the model FSS, is discussed in Section 3.
Hyperon s.p. potentials in nuclear matter are discussed
in Section 4. Summary is drawn in Section 5.

\section{$G$-matrices of the quark-model potential}

The quark-model interaction is defined through the
RGM equation for the parity-projected relative
wave function $\chi_\alpha(\bR)$ of the $(3q)$-$(3q)$ clusters:
\begin{equation}
\left[~\varepsilon_\alpha + {\hbar^2 \over 2\mu_\alpha}
\left( {\partial \over \partial \bR} \right)^2~\right]
\chi_\alpha(\bR)=\sum_{\alpha^\prime} \int d \bR^\prime
~G_{\alpha, \alpha^\prime}(\bR, \bR^\prime; E)
~\chi_\alpha(\bR^\prime)\ \ .
\label{fm1}
\end{equation}
The subscript $\alpha$ specifies a set of quantum
numbers of the channel wave function, $\alpha=
\left[1/2(11)\,a_1, 1/2(11)a_2 \right]$ $SS_zYII_z; \CP$,
where $\CP$ is the flavor-exchange phase
and $1/2(11) a$ denotes the spin, the $SU_3$ quantum number
in the Elliott notation $(\lambda \mu)$,
and the flavor label $a=YI$ of the octet
baryons, respectively.
For example, $YI=1(1/2)$ for $N$, $00$ for $\Lambda$,
and $01$ for $\Sigma$.
The relative energy $\varepsilon_\alpha$ of the
channel $\alpha$ is related to the total energy $E$ of the system
through $\varepsilon_\alpha=E-E^{int}_{a_1}
-E^{int}_{a_2}$ with $E^{int}_{a_i}$ being
the intrinsic energy of the baryon.
The exchange kernel $G_{\alpha \alpha^\prime}(\bR, \bR^\prime; E)$
is given by
\begin{eqnarray}
& & G_{\alpha \alpha^\prime}(\bR, \bR^\prime; E)
= \delta(\bR-\bR^\prime) \left[
~\sum_\beta V^{(CN)\beta}_{\alpha \alpha '\,D}(\bR)
+\sum_\beta V^{(SN)\beta}_{\alpha \alpha '\,D}(\bR) \right. \nonumber \\
& & \left. +\sum_\beta V^{(TN)\beta}_{\alpha \alpha '\,D}
(\bR)~(S_{12})_{\alpha \alpha '}~\right] + \sum_\Omega
\CM^{(\Omega)}_\aagp (\bR, \bR^\prime)
-\varepsilon_\alpha~\CM^N_{\alpha \alpha^\prime}(\bR, \bR^\prime)
\ \ .
\label{fm2}
\end{eqnarray}
The quark exchange kernel $\CM^{(\Omega)}_\aagp
(\bR, \bR^\prime)$ on the right hand side of \eq{fm2}
includes a sum over $\Omega=K$ for the exchange kinetic-energy kernel,
various pieces of the Fermi-Breit interaction, as well as
$CN$ for the central term of the scalar- (S-) meson exchange,
$SN$ for the spin-spin term of the pseudo-scalar-
(PS-) meson exchange, and $TN$ for the tensor term
of the PS-meson exchange.
For details, \cite{NA95} and \cite{FU96b} should be referred to.
According to \cite{FU97}, we introduce the basic Born
kernel of \eq{fm2} through
\begin{equation}
M_\aagp (\bq_f, \bq_i; E)
=\langle\,e^{i \bq_f \cdot \bR}\,\vert\,G_\aagp (\bR, \bR^\prime; E)
\,\vert\,e^{i \bq_{\,i}\cdot {\bR}^\prime} \rangle \ \ .
\label{fm3}
\end{equation}
The full Born kernel of the quark-exchange kernel is given by
\begin{equation}
V_{\gamma \alpha}(\bp, \bq; E)={1 \over 2}
\left[ M_{\gamma \alpha}(\bp, \bq; E)
+(-1)^{S_\alpha} {\CP}_\alpha M_{\gamma \alpha}
(\bp, -\bq; E) \right]\ \ .
\label{fm4}
\end{equation}
It is now straightforward to convert the RGM equation \eq{fm1}
to the Lippmann-Schwinger-type equation \cite{FU99}, which takes the form
\begin{eqnarray}
& & T_{\gamma \alpha}(\bp, \bq; E)
=V_{\gamma \alpha}(\bp, \bq; E)
+\sum_\beta {1 \over (2\pi)^3} \int d \bk
~V_{\gamma \beta}(\bp, \bk; E) \nonumber \\
& & \times {2 \mu_\beta \over \hbar^2}
{1 \over k_\beta^2-k^2+i \varepsilon}
~T_{\beta \alpha}(\bk, \bq; E)\ \ .
\label{fm5}
\end{eqnarray}
Here the total energy $E$ is related to $k_\beta^2$ through
$E=\varepsilon_\beta+E^{int}_{b_1}+E^{int}_{b_2}$
with $\varepsilon_\beta=(\hbar^2 k_\beta^2/2\mu_\beta)$.

The Bethe-Goldstone (BG) equation for the $G$-matrix solution
is obtained by replacing the propagator in \eq{fm5} as
\begin{equation}
{2 \mu_\beta \over \hbar^2}
{1 \over k_\beta^2-k^2+i \varepsilon}
\rightarrow {Q_\beta(k, K) \over e_\beta(k, K; \omega)}\ \ ,
\label{fm6}
\end{equation}
where $Q_\beta(k, K)$ stands for the angle-averaged Pauli operator
and $e_\beta(k, K; \omega)$ is the energy denominator
given by\footnote{For s.p. potentials, we use the notation
$b_1 =b$ and $a_1 =a$ to specify baryons.}
\begin{equation}
e_\beta(k, K; \omega)=\omega-E_{b}(k_1)-E_N(k_2)\ \ .
\label{fm7}
\end{equation}
Explicit expressions for $Q_\beta$ and $k_i$ are discussed below.
First $E_b(k)$ is the s.p. energy defined through
\begin{equation}
E_b(k)=M_b+{\hbar^2 \over 2M_b} k^2+U_b(k)\ \ ,
\label{fm8}
\end{equation}
with $U_b(k)$ and $M_b$ being the s.p. potential
and the mass for the baryon $b$, respectively.\footnote{It should be
noted that $M_b=E_b^{int}$ since the threshold energy is
usually fitted to the experimental value in our model.}
The starting energy $\omega$ is a sum
of the s.p. energies of two interacting baryons:
\begin{eqnarray}
& & \omega = E_{a}(k_1)+E_N(k_2) \nonumber \\
& & = M_a + M_N +{\hbar^2 \over 2(M_{a}+M_N)} K^2
+{\hbar^2 \over 2 \mu_\alpha} q^2
+U_a(k_1)+U_N(k_2)\ \ ,
\label{fm9}
\end{eqnarray}
where $\bK$ and $\bq$ are the total and relative momenta
corresponding to the initial s.p. momenta $\bk_1$ and $\bk_2$.
The BG equation is, therefore, expressed as
\begin{eqnarray}
G_{\gamma \alpha}(\bp, \bq; \omega)
& = & V_{\gamma \alpha}(\bp, \bq; E)
+\sum_\beta {1 \over (2\pi)^3} \int d~\bk
~V_{\gamma \beta}(\bp, \bk; E) \nonumber \\
& & \times {Q_\beta(k, K) \over e_\beta(k, K; \omega)}
~G_{\beta \alpha}(\bk, \bq; \omega)\ \ ,
\label{fm10}
\end{eqnarray}
and the s.p. potential is calculated from
\begin{eqnarray}
& & U_a(k_1) = \sum_{\sigma_2, \tau_2} \int_{|\bk_2| < k_F}
d~\bk_2~\langle~a\,\bk_1~N\, \bk_2~\vert \nonumber \\
& & \times G(\omega=E_a (k_1)+E_N (k_2))
~\vert~a\,\bk_1~N\,\bk_2 - N\,\bk_2~a\,\bk_1~\rangle \ \ ,
\label{fm11}
\end{eqnarray}
with $k_F$ being the Fermi-momentum of symmetric nuclear
matter.
In \eq{fm11} the sum over $\sigma_2, \tau_2$ implies
the spin-isospin sum with respect to the nucleons
in nuclear matter.

There exists an inherent ambiguity of how to deal with the energy
dependence of the Born kernel $V_{\gamma \alpha}(\bp, \bq; E)$ in
the BG equation \eq{fm10}. The total energy of the two
interacting particles in the nuclear medium is not conserved.
Since we only need the diagonal $G$-matrices for calculating
s.p. potentials, we here simply use
\begin{equation}
\varepsilon_\gamma=E_a^{int} -E_c^{int}
+{\hbar^2 \over 2\mu_\alpha}q^2\ \ ,
\label{fm12}
\end{equation}
both in $V_{\gamma \alpha}(\bp, \bq; E)$ and $V_{\gamma \beta}
(\bp, \bk; E)$ in \eq{fm10}.

We use two different assumptions for the s.p. potentials
involved in $E_{b}(k_1)+E_N(k_2)$ term of \eq{fm7}.
The first one is the so-called QTQ prescription defined
by taking only the rest mass and kinetic-energy term
in Eqs.\,(\ref{fm7}) and (\ref{fm8}).
The other case, the continuous choice, is given by the s.p.
potential calculated self-consistently.
If the energy denominator $e_\beta(k, K; \omega)$ involves
a pole, the solution of the BG equation becomes complex,
so does the s.p. potential $U_a(q_1)$ in \eq{fm11}.
In this case, only the real part Re $U_b(k)$ is retained
in $e_\beta(k, K; \omega)$.
As for the Pauli operator $Q_\beta(k, K)$,
we adopt the standard angle-average approximation. Denoting
the mass, the momentum and the Fermi momentum of the one particle
by $M_1$, {\boldmath $k$}$_1$ and $k_F^{(1)}$ and those of the
other by $M_2$, {\boldmath $k$}$_2$ and $k_F^{(2)}$,
the angle-averaged Pauli operator for the total and relative
momenta of {\boldmath $K$}={\boldmath $k$}$_1$+{\boldmath $k$}$_2$
and $\bk=(\xi \bk_1-\bk_2)/(1+\xi)$ with $\xi=(M_2/M_1)$ is
given by
\begin{eqnarray}
& & Q_\beta(k, K)={1 \over 2}\int_{-1}^{1}
\;d\cos\theta\; \nonumber \\
& & \times \Theta \left(\,\left|{1 \over1+\xi}
\mbox{{\boldmath $K$}}+\mbox{{\boldmath $k$}}\right| -k_F^{(1)}\right)
~\Theta \left(\,\left| {\xi \over 1+\xi}
\mbox{{\boldmath $K$}}-\mbox{{\boldmath $k$}}\right| -k_F^{(2)}\right),
\label{fm13}
\end{eqnarray}
where $\Theta$ is the Heaviside's step function and $\theta$ denotes
the angle between {\boldmath $K$} and {\boldmath $k$}.
This becomes
\begin{eqnarray}
& & Q_\beta(k, K)=[0|\left([-1|z_1 |1]
+[-1 |z_2|1]\right)/2|1]\ \ , \nonumber \\
& & z_1={ 1 \over 2} \left(1+\xi \right){1 \over kK}
\left\{ \left( {1 \over 1+\xi}K\right)^2 + k^2 -(k_F^{(1)})^2 \right\}\ \ ,
\nonumber \\
& & z_2={1 \over 2}\left(1+{ 1 \over \xi}\right)
{1 \over kK} \left\{ \left({\xi \over 1+\xi}K\right)^2 + k^2 -(k_F^{(2)})^2
\right\}\ \ ,
\label{fm14}
\end{eqnarray}
where we have introduced
the notation $[a|b|c] \equiv \max (a, \hbox{min} (b,c))$
as is used in \cite{SCHU}.
For the $YN$ system, we set $k_F^{(1)}=0$ and $k_F^{(2)}=k_F$,
which results in $Q_\beta(k,K)=(1+[-1|z_0|1])/2$ with
\begin{equation}
z_0=\frac{1}{2} \left( 1+\frac{1}{\xi}\right){1 \over kK}
\left[\left(\frac{\xi}{1+\xi}K \right)^2 +k^2 - k_F^2\right]\ \ .
\label{fm15}
\end{equation}
For the $NN$ system,
$k_F^{(1)}=k_F^{(2)}=k_F$ yields $Q_\beta(k,K)=[0|z_0|1]$ with $\xi =1$.

It is convenient to deal with partial-wave components of the
Born kernel, by carrying out angular integration numerically.
We define $V^{J}_{\gamma S^\prime \ell^\prime,
\alpha S \ell}(p, q; E)$ through
\begin{eqnarray}
& & V_{\gamma \alpha}(\bp, \bq; E)
=\sum_{J M \ell \ell^\prime}^{\qquad \prime}
4\pi~V^{J}_{\gamma S^\prime
\ell^\prime, \alpha S \ell}(p, q; E) \nonumber \\
& & \times \sum_{m^\prime} \langle \ell^\prime m^\prime
S^\prime S^\prime_z \vert JM \rangle~Y_{\ell^\prime m^\prime}
(\widehat{\bp})
~\sum_m \langle \ell m S S_z \vert JM \rangle~Y^*_{\ell m}
(\widehat{\bq})\ \ ,
\label{fm16}
\end{eqnarray}
where $\gamma=\left[1/2(11)\,c_1, 1/2(11)c_2 \right]$ $S^\prime
S^\prime_z Y I I_z; \CP^\prime$, and the prime
in $\sum^\prime$ indicates
that the sum over $\ell$ and $\ell^\prime$ is only
for $(-1)^{S^\prime}\CP^\prime=(-1)^{\ell^\prime}
=(-1)^S\CP=(-1)^\ell= \hbox{parity}$.
Then it is straightforward to apply \eq{fm16} to
the partial-wave decomposition \cite{HT70} of the BG equation
in nuclear matter:
\begin{eqnarray}
& & G^J_{\gamma S^\prime \ell^\prime, \alpha S \ell}(p, q; K, \omega)
=V^{J}_{\gamma S^\prime \ell^\prime, \alpha S \ell}(p, q; E)
+{4\pi \over (2\pi)^3} \sum_{\beta S^{\prime \prime}
\ell^{\prime \prime}}^{\qquad \prime}
 \int^{\infty}_{0} k^2\,d\,k \nonumber \\
& & \times V^{J}_{\gamma S^\prime \ell^\prime, \beta S^{\prime \prime}
\ell^{\prime \prime}}(p, k; E)
~{Q_\beta(k, K) \over e_\beta(k, K; \omega)}
~G^J_{\beta S^{\prime \prime} \ell^{\prime \prime}, \alpha S \ell}
(k, q; K, \omega)\ \ .
\label{fm17}
\end{eqnarray}
The s.p. potential is calculated from
\begin{eqnarray}
& & U_a(k_1)=(1+\delta_{a,N}) \sum_I
{2I+1 \over 2(2I_a+1)} \nonumber \\
& & \times \sum_{J\ell S} (2J+1){1 \over (2\pi)^2}
\int_{-1}^1 \;d\cos \theta_2
\int_{0}^{k_F} k_2^2\,d\,k_2
~G_{aS\ell, aS\ell}^J (q, q; K, \omega)\ \ ,
\label{fm18}
\end{eqnarray}
where $q$, $K$ and $\omega$ are given by
\begin{eqnarray}
& & q=\frac{1}{1 + \xi}\left[ \xi^2 k_1^2+k_2^2
- 2\xi k_1 k_2 \cos\theta_2
~\right]^{{1 \over 2}}\ \ ,
\qquad \xi={M_N \over M_a}\ \ ,\nonumber \\
& & K= \left[k_1^2+k_2^2
 + 2 k_1 k_2 \cos \theta_2~\right]^{{1 \over 2}}\ \ ,
 \nonumber \\
& & \omega=E_a(k_1)+E_N(k_2)\ \ .
\label{fm18b}
\end{eqnarray}

If we average the $K$-dependence of the $G$-matrix, the calculation
can be further simplified. Changing the integral variable
to the relative momentum $\bq$, the expression for the $U_a (k_1)$
becomes
\begin{eqnarray}
& & U_a(k_1)=(1+\delta_{a,N})(1+\xi)^3 \sum_I
{2I+1 \over 2(2I_a+1)} \nonumber \\
& & \times \sum_{J\ell S} (2J+1) {1 \over 2 \pi^2}
\int_{0}^{q_{max}} q^2\,d\,q
~W(k_1, q)~G_{aS\ell, aS\ell}^J (q, q; K, \omega)\ \ ,
\label{fm18c}
\end{eqnarray}
where $q_{max}=(k_F+\xi k_1)/(1+\xi)$,
and $W(k_1, q)$ is the phase space factor given by
\begin{equation}
W(k_1, q)={1 \over 2}\left(1-[-1|x_0|1]\right)~~\hbox{with}
~~x_0={\xi^2 k_1^2+(1+\xi)^2 q^2-k_F^2 \over 2 \xi (1+\xi) k_1 q}\ .
\label{fm19}
\end{equation}
Once $k_1$ and $q$ are given, the values of $K$, $k_2$ and $\omega$ are
calculated through
\begin{eqnarray}
& & K=(1+\xi) \left[k_1^2+q^2-k_1 q \left( 1+[-1|x_0|1] \right)
~\right]^{{1 \over 2}}\ \ , \nonumber \\
& & k_2=\left[ {\xi \over 1+\xi} K^2+(1+\xi) q^2 - \xi k_1^2
~\right]^{{1 \over 2}}\ \ ,\nonumber \\
& & \omega=E_a(k_1)+E_N(k_2)\ \ .
\label{fm20}
\end{eqnarray}
In Eqs.\,(\ref{fm19}) and (\ref{fm20}), we should assume $x_0=-1$
when $k_1=0$.

In the continuous prescription, we need to determine
the s.p. momenta $k_1$ and $k_2$ in \eq{fm7} for the intermediate
energy spectra. These are given through
\begin{eqnarray}
& & {k_1}^2=\left({1 \over 1+\xi}K\right)^2+k^2
+{1 \over 1+\xi}kK\left([-1|z_2|1]-[-1|z_1|1]\right)\ \ ,
\nonumber \\
& & {k_2}^2=\left({\xi \over 1+\xi}K\right)^2+k^2
-{\xi \over 1+\xi}kK\left([-1|z_2|1]-[-1|z_1|1]\right)\ \ ,
\label{fm21}
\end{eqnarray}
in the two Fermi-sphere case.
For the $NN$ system, there is a special situation that
the linear term of $z$ vanishes in the process of
angular averaging. In this case, we use the angle-average
over $z^2$, as is discussed in \cite{SCHU}, and use
\begin{eqnarray}
& & {k_1}^2={1 \over 4}K^2+k^2
+kK{1 \over \sqrt{3}}[0|z_0|1]\ \ ,\nonumber \\
& & {k_2}^2={1 \over 4}K^2+k^2
-kK{1 \over \sqrt{3}}[0|z_0|1]\ \ ,
\label{fm22}
\end{eqnarray}
where $z_0$ is given by \eq{fm15} with $\xi =1$.

Finally, the ground-state energy per nucleon of symmetric nuclear
matter is given by
\begin{equation}
 {E \over A}={3 \over 5}\left(\frac{\hbar^2}{2M_N} {k_F}^2 \right)
+{3 \over 2k_F^3} \int_{0}^{k_F} k^2\,d\,k \;U_N (k)\ \ .\nonumber \\
\label{fm23}
\end{equation}
%
%
%
%
%

\begin{figure}[t]
\epsfxsize=7cm
\hspace*{2.9cm}
\epsfbox{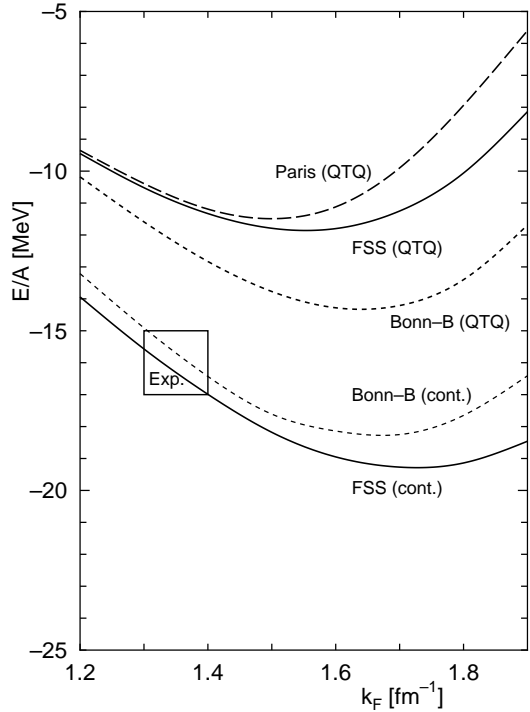}
\caption{Nuclear matter saturation curves obtained for the quark-model
potential FSS, together with the results with the Paris potential
\protect\cite{PARI} and the Bonn potential \protect\cite{BONN}.
The choice of the intermediate spectra is specified by "QTQ" and "cont.",
respectively. The result for the Bonn-B potential with the continuous
choice is taken from the nonrelativistic calculation
in \protect\cite{BM}.}
\end{figure}

\section{Saturation properties of nuclear matter}

Figure 1 shows saturation curves calculated for ordinary nuclear matter
with the QTQ prescription as well as the continuous choice for
intermediate spectra,
together with the results of the Paris potential \cite{PARI}
and the Bonn B potential \cite{BONN}.
The $k$-dependence of the nucleon s.p. potential $U_N (k)$ obtained
with the continuous choice is shown in Fig.\,2 at three densities
$\rho = 0.5 \rho_0$, $0.7\rho_0$ and $\rho_0$ with the normal
density $\rho_0$ = 0.17 fm$^{-3}$, which correspond to
$k_F =1.07,\;1.2,\;1.35$ fm$^{-1}$.
For comparison, the result with
the Nijmegen soft-core (NSC) potential NSC89 \cite{NSC} calculated
by Schulze {\it et al.} \cite{SCHU} is also shown.
At lower momentum, $k < 2k_F$, our result is close to the potential
obtained with the NSC $NN$ potential.

In the calculation
with the continuous prescription, it turns
out that the nucleon s.p. potential starts to decrease at
$k \sim 4$ fm$^{-1}$, the tendency of which is seen in Fig.\,2.
This behavior of the s.p. potential suggests
that the short-range repulsion of the $NN$ interaction in the FSS
may not be strong enough at higher energies, namely beyond the region
where the parameters are fixed to reproduce the scattering data.
Actually, this is not the case since the $NN$ differential
cross sections are not largely overestimated at higher energy region
even up to 800 MeV \cite{HYP97}.
Detailed analysis \cite{FU99} implies that
this particular feature of FSS is caused
by the ill-behavior of the spin-independent central
invariant amplitude at the forward angles, which is intimately
related to the asymptotic behavior of s.p. potentials in the
high-momentum region.
This flaw of the model FSS may be removed by introducing
the momentum-dependent higher-order term involved in the S-meson
exchange central force.
In the present application, we introduced an ad hoc prescription
to set $U_N (k)=U_N (k=3.8~\hbox{fm}^{-1})$ for $k \geq 3.8
~\hbox{fm}^{-1}$ in case of the continuous choice
for the intermediate spectra.\footnote{The value
of $k=3.8~\hbox{fm}^{-1}$ corresponds to the
incident energy $T_{lab}=300~\hbox{MeV}$ in the two-nucleon
scattering.}
Since the kinetic energy term dominates at higher energies,
the result does not depend much on this particular prescription.

The short-range part of the FSS is mainly described by the quark-exchange
mechanism. The non-local character of this part is different from
the usual vector meson exchange picture in the one-boson exchange model.
In spite of the difference the saturation point of the quark-model
FSS is seen not to deviate from the Coester band,
which indicates that the FSS has similar saturation properties
with other realistic meson-exchange potentials.

\begin{figure}[t]
\epsfxsize=7.5cm
\hspace*{2.5cm}
\epsfbox{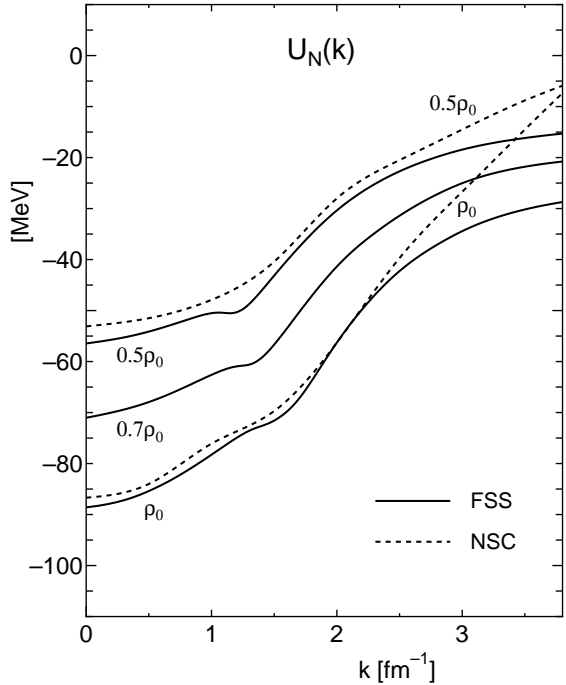}
\caption{The nucleon s.p. potential \(U_N (k)\) in nuclear
matter with the continuous choice for intermediate spectra.
The quark model is FSS. Potentials for three densities
of \(\rho= 0.5 \rho_0,\,0.7 \rho_0,\,\rho_0\) are shown, where
the normal density \(\rho_0\) corresponds to \(k_F =1.35\) fm\(^{-1}\).
The dashed curve is the result by Schulze {\it et al.}
\protect\cite{SCHU} with
the Nijmegen soft-core $NN$ potential NSC89 \protect\cite{NSC}.}
\end{figure}
%

\section{Hyperon potentials in nuclear matter}

For hyperon s.p. potentials in nuclear matter,
we show the results with the continuous prescription for intermediate
spectra. Since the $\Lambda$ and $\Sigma$ are coupled
in the isospin $I=1/2$ channel, these potentials are calculated
self-consistently. Since the asymptotic behavior in the high-momentum
region seems to be unsatisfactorily described
as in the case of the $NN$ channel,
the same prescription in the energy denominator
as for the nucleon is introduced; namely,
$U_{\Lambda} (k)=U_{\Lambda} (k=3.8 \mbox{ fm}^{-1})$ and
$U_{\Sigma} (k)=U_{\Sigma} (k=3.8 \mbox{ fm}^{-1})$
for $k \geq 3.8~\mbox{fm}^{-1}$.

\subsection{$\Lambda$ s.p. potential}

Figs.\,3 shows the momentum dependence of the \(\Lambda\)
s.p. potential in nuclear matter obtained
from quark-model $G$-matrices. For comparison, the result by
Schulze {\it et al.} \cite{SCHU} with
the NSC $YN$ potential NSC89 \cite{NSC} is also shown.
Partial wave contributions
of the s.p. potential \(U_{\Lambda} (k=0)\) in nuclear matter
at $k_F = 1.35~\hbox{fm}^{-1}$ are tabulated
in Table 1.

\begin{figure}[b]
\epsfxsize=7.5cm
\hspace*{2.5cm}
\epsfbox{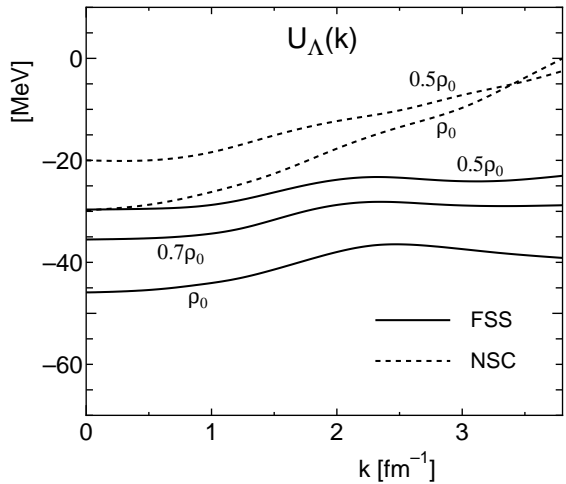}
\caption{The $\Lambda$ s.p. potential \(U_{\Lambda} (k)\) in symmetric
nuclear matter with the continuous choice for intermediate spectra.
The quark model is FSS. Potentials for three densities
of \(\rho= 0.5 \rho_0,\,0.7 \rho_0,\,\rho_0\) are shown, where
the normal density \(\rho_0\) corresponds to \(k_F =1.35\) fm\(^{-1}\).
The dashed curve is the result by Schulze {\it et al.}
\protect\cite{SCHU} with
the Nijmegen soft-core $YN$ potential NSC89 \protect\cite{NSC}. \hfill
}
\end{figure}
\begin{table}[t]
\caption{\(\Lambda\) and \(\Sigma\) s.p. potentials
in nuclear matter with \(k_F = 1.35\) fm\(^{-1}\),
calculated from our quark-model (FSS) \(G\)-matrices
with the continuous prescription for intermediate spectra.
Predictions by Nijmegen soft-core potential (NSC) \protect\cite{NSC} is
also shown for comparison \protect\cite{SCHU}.
}
\label{table1}
\renewcommand{\arraystretch}{1.3}
\setlength{\tabcolsep}{3mm}
\begin{center}
\begin{tabular*}{\textwidth}{@{}c@{\extracolsep{\fill}}rrcrrrr} \hline
 & \multicolumn{2}{c}{$U_\Lambda(0)$ \hspace{1em}[MeV]} &
 & \multicolumn{4}{c}{$U_\Sigma(0)$ \hspace{1em}[MeV]} \\
\cline{2-3} \cline{5-8}
 & FSS & NSC & & \multicolumn{2}{c}{FSS}
 & \multicolumn{2}{c}{NSC} \\ \hline
$I$ & 1/2 & 1/2 &  & 1/2 & 3/2 & 1/2 & 3/2 \\ \hline
$\hbox{}^1S_0$ & $-19.9$ & $-15.3$ &  & $6.1$
& $-8.8$ & $6.7$ & $-12.0$ \\
$\hbox{}^3S_1+\hbox{}^3D_1$ & $-21.2$ & $-13.0$ &  & $-19.7$ & $48.3$
 & $-14.9$ & $6.7$\\
$\hbox{}^1P_1+\hbox{}^3P_1$ & $0.2$ & $3.6$ &  & $-6.7$ & $4.0$
 & $-3.5$ & $3.9$ \\
$\hbox{}^3P_0$ & $0.6$ & $0.2$ &  & $3.0$ & $-2.3$ & $2.6$
& $-2.0$ \\
$\hbox{}^3P_2+\hbox{}^3F_2$ & $-4.6$ & $-4.0$  &  & $-1.2$ & $-1.2$
& $-0.5$ & $-1.9$\\ \hline
 subtotal  &   &  & & $-21.0$ & $41.4$ & $-9.8$ & $-5.5$ \\
total & $-45.9$ & $-29.8$ & & \multicolumn{2}{c}{$+20.7$}
 & \multicolumn{2}{c}{$-15.3$} \\
\hline
\end{tabular*}
\end{center}
\end{table}

The FSS is seen to produce a larger potential
depth $U_{\Lambda}(0)$ than the NSC.
Actually the depth of 46 MeV in the
case of $k_F =1.35$ fm$^{-1}$ is larger than that of the
standard phenomenological potential depth of about 30 MeV \cite{BMZ}.
It is, however, necessary to consider the density dependence
of the potential in order to relate
$U_{\Lambda}(0)$ to the empirical potential
in finite nuclei. The correction from the starting energy dependence
also modifies the s.p. potential.

If we compare the FSS and NSC results in Table 1, we find that
the additional attraction of our $U_{\Lambda}(0)$ to the NSC
originates from three sources;
1) about 5 MeV excess for $\hbox{}^1S_0$ state,
2) about 8 MeV excess for $\hbox{}^3S_1+\hbox{}^3D_1$ state,
and 3) about 3 MeV excess for $\hbox{}^1P_1+\hbox{}^3P_1$ state.
The first one is apparently because our $\Lambda N$ $\hbox{}^1S_0$ state
is too attractive as is seen from the phase-shift behavior
shown in Fig.\,4 of \cite{FU96b}. The detailed analysis of
the $s$-shell $\Lambda$-hypernuclei seems to imply that
the relative strength of the attraction
of the $\hbox{}^1S_0$ and $\hbox{}^3S_1$ states is most
desirable to be such that the
maximum peak of the $\hbox{}^1S_0$ phase shift is larger
than that of $\hbox{}^3S_1$ by
about $10^\circ$ \cite{SH83,YA94,HI97}.
If we fit the available low-energy $\Lambda p$ total cross section
data, this condition yields $\delta_{max}(\hbox{}^1S_0) \sim 34^\circ
~\hbox{-}~36^\circ$ and $\delta_{max}(\hbox{}^3S_1)
\sim 26^\circ$ approximately. On the other hand, the FSS (and
also RGM-H) prediction is $\delta_{max}(\hbox{}^1S_0) \sim 46^\circ$
and $\delta_{max}(\hbox{}^3S_1) \sim 17^\circ$, and our $\hbox{}^1S_0$
state is too attractive. For the contribution 2),
it is a puzzle why these low values of the $\hbox{}^3S_1$ phase shift
give such a large attractive contribution as $-21.2$ MeV.
It can be checked by switching off the $\Lambda N-\Sigma N$
transition interaction that about a half of the attractive contribution
comes through the $\Lambda N-\Sigma N$ coupling.

As to the contribution 3), we should note that $P$-state interaction
of our quark model is weakly attractive in the low-energy region,
because of a very strong effect of the
antisymmetric $LS$ force ($LS^{(-)}$ force) \cite{FJ98}.
Among many versions of the Nijmegen models, only the model D gives
attractive $P$-wave interaction. However, the mechanism of producing
the attraction is entirely different between our model and the
model D. In our case, a very strong $\Lambda N$-$\Sigma N (I=1/2)$
coupling takes place around the $\Sigma N$ threshold region
in the $\hbox{}^1P_1+\hbox{}^3P_1$ state,
which is  due to the $LS^{(-)}$ force
originating from the Fermi-Breit spin-orbit interaction.
A broad $P$-state resonance exists either
in the $\Sigma N (I=1/2)$ $\hbox{}^3P_1$ state
or in the $\Lambda N$ $\hbox{}^1P_1$ state, depending on
the strengths of the coupling matrix elements and of the central
attraction in the $\Sigma N (I=1/2)$ channel.
The $\hbox{}^1P_1$ and $\hbox{}^3P_1$ phase shifts become
attractive even at $p_{lab}=300~\hbox{MeV}/c$ by the influence
of this resonance. This attractive behavior of the $\Lambda p$
$P$-wave interaction can be examined experimentally by observing
the forward-to-backward ratio of the $\Lambda p$ differential
cross sections in the $p_{lab} \leq
300~\hbox{MeV}/c$ region \cite{FJ98}.

Figure 3 also indicates that the momentum dependence
of the $\Lambda$ s.p. potential is weak. If we define a global
effective mass by
\begin{equation}
\frac{m^* (k)}{m} = \left[ 1+ \frac{2m}{\hbar^2 k^2}
(U(k)-U(0))\right]^{-1}
\label{fm26}
\end{equation}
$m^* (k\sim 1~\mbox{fm}^{-1})/m \sim$ 0.90, 0.94 and 0.95
at $k_F$ = 1.35, 1.2 and 1.07 fm$^{-1}$, respectively,
which is larger than that of the nucleon of 0.66, 0.72 and 0.79 at
the corresponding $k_F$. The potential from the NSC is also seen to
show a flat $k$-dependence similar to the FSS.
Supposing that the interaction $G$ in eq. (11) is a local
two-body potential, the direct term does not give $k$-dependence
and the exchange term, which is a strangeness exchange process,
is the origin of the effective mass. Although the two-body
correlation together with the Pauli operator $Q$ in the BG
equation brings about the momentum dependence even in the
direct term, the chief source of the effective mass in
the hyperon s.p. potential is considered to be the strangeness
exchange process. The possible reason for the difference of the lambda
and nucleon effective masses is the absence of the long-ranged
one-pion exchange in the $\Lambda$ case.

\begin{figure}[t]
\epsfxsize=7.5cm
\hspace*{2.5cm}
\epsfbox{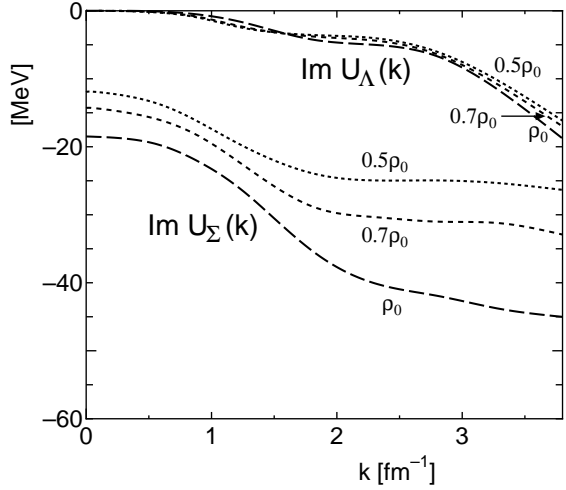}
\caption{The imaginary part of the $\Lambda$ and $\Sigma$ s.p.
potentials $U_a (k)$ in symmetric nuclear matter with the continuous
choice for intermediate spectra. The quark model is FSS.
Potentials for three nuclear matter densities
of \(\rho= 0.5\rho_0,\,0.7\rho_0,\,\rho_0\) are shown,
where the normal density \(\rho_0\) corresponds
to \(k_F =1.35\) fm\(^{-1}\).
}
\end{figure}

The imaginary part of the $\Lambda$ s.p. potential is shown in Fig. 4.
The imaginary part comes from creating
a nucleon one-particle-one-hole state. Since only the nucleons near
the Fermi surface participate for inelastic processes of the $\Lambda$'s
having small $k$, the imaginary strength is small for these lambda states,
which corresponds to the small spreading width of the $\Lambda$ formation
peaks observed in $(K, \pi)$ and $(\pi, K)$ reactions \cite{KPI}.

\subsection{$\Sigma$ s.p. potential}

The momentum dependence of the $\Sigma$ s.p. potential
is shown in Fig.\,5. The result by Schulze {\it et al.} \cite{SCHU}
with the NSC $YN$ potential NSC89 \cite{NSC} is also shown.
The $U_{\Sigma} (k=0)$ turns out to be positive, which
is a marked difference from the results of other $YN$ potentials
except for the Nijmegen model F.
Table 1 presents each partial wave contribution
to the $U_{\Sigma}(0)$, which shows the strong repulsive
contribution in $I=3/2$ ${^3}S_1 + {^3}D_1$ channel.
This repulsion is a direct result of the strong
quark-antisymmetrization effect
in this particular channel \cite{FU96b}.

\begin{figure}[b]
\epsfxsize=7.5cm
\hspace*{2.5cm}
\epsfbox{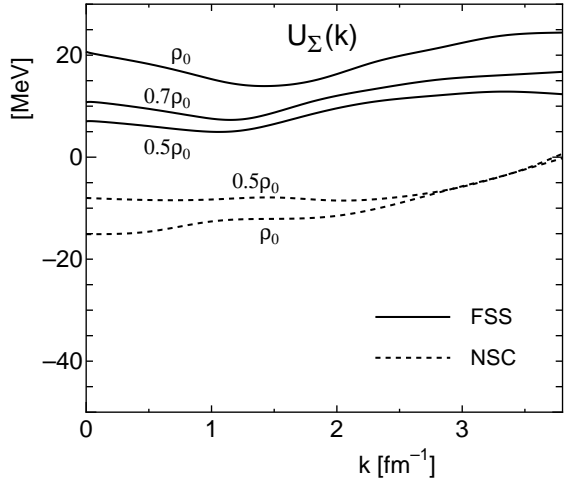}
\caption{The same as Fig.\,3 but
for the $\Sigma$ s.p. potential \(U_{\Sigma}(k)\). \hfill}
\end{figure}

Since the repulsive $\Sigma$ s.p. potential is not common
with many other realistic $YN$ potentials, it would be useful
to examine predictions by the other versions
of our quark model, RGM-F and RGM-H.
Because of a technical reason, it is not possible to obtain
a reasonable s.p. potential for $\Lambda$ in these models.
The completely Pauli-forbidden $(11)_s$ component
in the $\Lambda N$-$\Sigma N (I=1/2)$ coupled-channel system is
not exactly eliminated in the present calculation.
In the $\Lambda N$ scattering, this
causes an unrealistic $\Lambda N$-$\Sigma N$ coupling
in the $\hbox{}^1S_0$ state, which was remedied
in the previous variational calculation by modifying
the spin-flavor factors for this particular channel \cite{FU95}.
It is, however, not easy to incorporate this prescription
in the Lippmann-Schwinger and $G$-matrix equations,
since these equations are formulated in the momentum representation.
Without this prescription, the inaccuracy
of the $\Lambda N$ $\hbox{}^1S_0$ phase shift sometimes
reaches at more than 10 degrees.
This is because the antisymmetrization of the effective-meson
exchange potentials is not exactly
treated in these models.\footnote{The same problem exists even
in the model FSS with respect to the exchange kinetic-energy kernel.
However, the effect to the phase shift is very small
and the phase-shift difference is less than 2 degrees.}
In spite of these difficulties, we believe that the $\Sigma$ s.p.
potentials are rather correctly calculated, since the phase-shift
behavior of the $\Sigma N$ channel is fairly similar between
the two cases with this prescription
and without this prescription.

\begin{table}[b]
\caption{$\Sigma$ s.p. potentials in nuclear matter
at $k_F = 1.35~\hbox{fm}^{-1}$, calculated from our three versions
of the quark model; RGM-H \protect\cite{FU95},
FSS \protect\cite{FU96a,FU96b},
and RGM-H \protect\cite{FU96b}.
The completely self-consistent s.p. potentials are used
with the continuous prescription for intermediate spectra,
together with \protect\eq{fm18c}.}
\label{table2}
\renewcommand{\arraystretch}{1.3}
\setlength{\tabcolsep}{3mm}
\begin{center}
\begin{tabular*}{\textwidth}{crrcrrcrr}
\hline
 & \multicolumn{8}{c}{$U_\Sigma(0)$ \hspace{1em}[MeV]} \\
\cline{2-9}
 & \multicolumn{2}{c}{RGM-F} & & \multicolumn{2}{c}{FSS}
 & & \multicolumn{2}{c}{RGM-H} \\
 \hline
$I$ & 1/2 & 3/2 & & 1/2 & 3/2 & & 1/2 & 3/2 \\
\hline
$\hbox{}^1S_0$ & $5.4$ & $-10.7$ & & $6.0$ & $-8.8$ & &
$9.3$ & $-13.6$ \\
$\hbox{}^3S_1+\hbox{}^3D_1$ & $-25.4$ & $43.5$ & &
$-20.7$ & $47.5$ & & $-17.9$ & $39.1$ \\
$\hbox{}^1P_1+\hbox{}^3P_1$ & $-10.8$ & $4.8$ & &
$-7.7$ & $3.9$ & & $-8.1$ & $1.6$ \\
$\hbox{}^3P_0$ & $3.2$ & $-3.3$ & &
$2.9$ & $-2.3$ & & $2.7$ & $-2.4$ \\
$\hbox{}^3P_2+\hbox{}^3F_2$ & $-2.1$ & $-5.0$ & &
$-1.3$ & $-1.2$ & & $-1.7$ & $-3.8$ \\
\hline
subtotal & $-31.8$ & $30.0$ & & $-22.3$ & $40.0$ & &
$-17.5$ & $21.0$ \\
total & \multicolumn{2}{c}{$-1.9$} & &
\multicolumn{2}{c}{$+17.6$} & &
\multicolumn{2}{c}{$+3.5$} \\
\hline
\end{tabular*}
\end{center}
\end{table}

Table 2 shows partial-wave contributions of $U_{\Sigma}(0)$
in nuclear matter at $k_F = 1.35~\hbox{fm}^{-1}$,
predicted by our three versions RGM-F, FSS and RGM-H.
Here we exceptionally used completely self-consistent
s.p. potentials $U_B(k)$ ($B=N,~\Lambda,~\Sigma$),
without introducing any cut-off at $k=3.8~\hbox{fm}^{-1}$,
for the intermediate spectra in the continuous
prescription.
We also employed the approximate angular integration
in \eq{fm18c}.
We find that the difference of the FSS prediction in Table 1 and
Table 2 is very small.
It is a common feature for the three versions
that they give similarly strong repulsion
in the $\Sigma N (I=3/2)$ $\hbox{}^3S_1 + {^3}D_1$ state.
It may look strange that $\Sigma N (I=1/2)$ $\hbox{}^1S_0$ state
gives only weak repulsion in all these three versions,
in spite of the fact that the phase-shift behavior
of this channel is almost comparably repulsive
to the $\Sigma N (I=3/2)$ ${^3}S_1 + {^3}D_1$ channel.
On the other hand, the attractive behavior
of the $\Sigma N (I=3/2)$ $\hbox{}^1S_0$
and $\Sigma N (I=1/2)$ $\hbox{}^3S_1 + {^3}D_1$ channels
reflects the characteristics of each version very well.
Namely, the low-energy behavior of the $\Sigma N (I=3/2)$
$\hbox{}^1S_0$ phase shift in RGM-H is too attractive,
corresponding to the slight over-estimation of the $\Sigma^+ p$ total
elastic cross sections at $p_{lab} \leq 300~\hbox{MeV}/c$
(see Fig.\,11(a) of \cite{FU96b}).
In the $\Sigma N (I=1/2)$ $\hbox{}^3S_1 + \hbox{}^3D_1$ channel,
it is discussed in \cite{FU99} and \cite{FU98} that
the central attraction becomes weaker for RGM-F,
FSS and RGM-H in this order.
As a measure of this strength, we calculated
the potential depth $V_{\Sigma N(I=1/2)}^C(\hbox{}^3S)$ for
the $\hbox{}^3S$ state through $\bp=0$ Wigner transform
of the non-local exchange kernel.
We find $V_{\Sigma N(I=1/2)}^C(\hbox{}^3S)=-38~\hbox{MeV}$,
$-24~\hbox{MeV}$, and $-18~\hbox{MeV}$ for RGM-F, FSS,
and RGM-H, respectively.
This corresponds to the contributions
from the $\Sigma N (I=1/2)$ $\hbox{}^3S_1+\hbox{}^3D_1$ channel
in Table 2 very well; namely,
$-24.5~\hbox{MeV}$, $-20.7~\hbox{MeV}$,
and $-17.9~\hbox{MeV}$ for RGM-F, FSS, and RGM-H, respectively.
For the $\Sigma N (I=1/2)$ $\hbox{}^1P_1+\hbox{}^3P_1$ contribution,
our quark model yields fairly large attraction of $-7 \sim -11$ MeV,
in comparison with the meson-exchange potentials.
This is again by the strong effect of the $LS^{(-)}$ force
in our quark model.
Tables 1 and 2 show that the total strength
of the $\Sigma$ s.p. potential in our quark model
has rather strong model dependence from $-2$ MeV to 20 MeV,
as a cancellation of the repulsive $I=3/2$ contribution
and the attractive $I=1/2$ contribution.
It seems to be very important to determine the isospin dependence
of the $\Sigma$ s.p. potential, as is suggested in \cite{DA99}.

The analysis \cite{BAT79} of the energy shifts and the widths
of $\Sigma^-$ atomic levels
in 1970's concluded that the $\Sigma$ s.p. potential
is attractive $-\mbox{Re}~V_{opt}^{\Sigma}(0)
\sim 25\,\hbox{-}\,30$ MeV.
The DWIA analysis \cite{KHSW} of the pion inclusive
spectra, which is related to the $\Sigma$-formation
in (K$^-$, $\pi^+$) reaction, suggests that
the $\Sigma$ potential is much weaker. The $G$-matrix calculation
with the NSC89 potential, using a continuous choice of s.p. potentials,
predicts the depth of about 15 MeV \cite{SCHU}. The calculations
in \cite{YB85} and \cite{YB90} indicate\footnote{A QTQ choice
with introducing the constant shift in the
energy denominator was used in these
calculations \cite{YB85} and \cite{YB90}.}
that the Nijmegen model D gives a similar
attractive potential.
On the other hand, the Nijmegen model F seems to predict a
repulsive s.p. potential.
The origin of 5.8 MeV repulsion, reported in \cite{YB90} for
the model F is again
the $\Sigma N (I=3/2)$ $\hbox{}^3S_1+\hbox{}^3D_1$ channel,
which gives a strongly repulsive contribution of 47.1 MeV,
comparable to our value 48.3 MeV in Table 1.
An attempt \cite{RMF} to extend
a relativistic mean-field theory to $\Sigma$-hypernuclei
predicted almost the same s.p. energy
for the $\Lambda$ and $\Sigma$, indicating
that the $\Sigma$ potential is as attractive as the $\Lambda$ one.
In the relativistic mean field description, calculated results depend on
the ratio of the coupling constants $\alpha_i \equiv
(g_{i \Sigma^0}/{g_{i N}})$ ($i= \sigma, \omega, \rho$).
The quark-meson coupling model \cite{QMC} also gave similar results.

In the mean time, the $\Sigma^-$ atomic data were reanalyzed \cite{BAT},
allowing more general density-dependence. This study indicates that the
$\Sigma$ potential is repulsive inside the nucleus.
Guided by this observation, it was shown \cite{MARE} that
the tuning of the ratio of the coupling
constants $\alpha_i$ can produce the
repulsive $\Sigma$ potential. Dabrowski \cite{DA99} showed
that the recent $(K^- ,\pi^{\pm})$ experiments
at BNL \cite{BNL} favor the repulsive $\Sigma$ potential,
the strength of which is about 20 MeV.
It is interesting to see that our FSS quark-model result is
in line with these recent observations. However, it should be
noted that as in the case of the $\Lambda$ particle the potential
at $k_F = 1.35$ fm$^{-1}$ may not be directly
related to the empirical s.p. potential.
The density and starting-energy dependences must be
taken into account.

In early pioneering stage of the sigma-hypernuclear
formation experiments \cite{BERT}, the observed spectra hinted narrow
peak structure for the $\Sigma$-formation even if they are in unbound
energy region.
Thus the calculation of the $\Sigma$ width in nuclear medium
was paid much attention. In recent experiments \cite{BNL}, however,
no narrow paeks were reported. In our calculation,
the $\Sigma$ s.p. potential is repulsive and
the $\Sigma$ bound state is unlikely except for the specific
light nuclei, where the spin-isospin dependence can produce $\Sigma$
bound states like the $_{\Sigma}^4$He \cite{HARA}.
If the $\Sigma$ s.p. potential is repulsive,
the strength of the imaginary part of the s.p. potential
is not related to the spreading width of the $\Sigma$ formation peaks.
The $k$-dependence of the imaginary $\Sigma$ s.p. potential is shown
in Fig. 4. The value of about 20 MeV near $k=0$
at $k_F = 1.35$ fm$^{-1}$ is in accord with the calculation
by Schulze {\it et al.} \cite{SCHU}.

\section{Summary}

We have presented the first attempt to apply the recent realistic
quark-model baryon-baryon interaction to nuclear matter
calculations.
The reaction matrices for the $NN$, $\Lambda N$ and $\Sigma N$
channels have been calculated in ordinary nuclear matter
by solving the Bethe-Goldstone equations
for the exchange kernel of the quark-model interaction.
We used mainly the interaction called FSS,
which was developed by Kyoto-Niigata
group \cite{FU96a,FU96b,FJ96a,FJ96b,FJ98}.
Since the quark model provides a unified framework to describe the
$NN$ and $YN$ interactions, it is very interesting to study the
predictions for hyperon properties in nuclear medium in this model.

The quark-model interaction is defined in formulating the RGM
equation for the relative wave function of the (3q)-(3q) clusters.
Thus we have to first define partial wave amplitudes in momentum space
by numerical angular integration of the quark exchange kernel. Then
self-consistent determination of the $NN$, $\Lambda N$ and $\Sigma N$
$G$-matrices in nuclear matter is straightforward.

In the nucleon sector the FSS gives similar saturation properties
for the nuclear matter as other realistic $NN$ potentials. This
result is probably non-trivial but interesting,
in view of the fact that the origin and the
description of the short-range part of the interaction is quite
different. The $\Lambda$ single-particle (s.p.) potential
has the depth of about $46$ MeV in the case of the continuous
prescription for intermediate energy spectra.
This value is slightly more attractive
than the value expected from the
experimental data of $\Lambda$-hypernuclei \cite{BMZ}.
The $\Sigma$ s.p. potential has turned out to be repulsive
with the strength of about 20 MeV,
the origin of which is traced back to the strong Pauli repulsion
in the $\Sigma N (I=3/2)$ $\hbox{}^3S_1$ state.
This result seems to be consistent
with the indication from the analysis by Dabrowski \cite{DA99} of
recent $(K^{-}, \pi^{\pm})$ experiments \cite{BNL} at BNL.
Future experiments will be expected to settle the problem
of the \(\Sigma\) s.p. potential.

Our calculation also indicates that the FSS is
not appropriate for predicting
the asymptotic behavior of s.p. potentials
in high-momentum region \cite{FU99}.
The present $NN$ two-body potential is not designed for
the application to such a high-energy region.
It would be interesting to examine how the
prediction of the model FSS is changed by introducing
momentum-dependent higher-order terms involved in the S-meson
exchange central force. The improvement
in this direction is now under way.

The $LS^{(-)}$ force is absent in the $NN$ interaction,
but it plays a characteristic role in the $YN$ interaction.
In fact, the quark-model interaction suggests an
important antisymmetric spin-orbit component. Thus the examination
of the combined effects of the $LS$ and $LS^{(-)}$ interactions
to the hyperon s.p. spin-orbit potential is very interesting.
In the present paper, however, we did not discuss this problem.
This subject is studied in a separate paper \cite{LS99}.

Finally we note that it will be an important future subject
to consider hyperonic nuclear matter in the scope of the
quark-model baryon-baryon interaction, since the study
of $\Lambda\Lambda$ and $\Xi N$ interactions
is also in progress \cite{NA97}.
Since the $\Sigma$ s.p. potential is repulsive
in the quark-model description,
the admixture of the $\Sigma$ particle is suppressed
and this, in turn, will affect the behavior
of the $\Lambda$ particles in dense hyperonic nuclear matter.

\end{document}